\documentclass[conference]{IEEEtran}
\IEEEoverridecommandlockouts
\usepackage{cite}
\usepackage{amsmath,amssymb,amsfonts}
\usepackage{algorithmic}
\usepackage{graphicx}
\usepackage{textcomp}
\usepackage{xcolor}
\usepackage{multirow}
\usepackage{tikz}
\usepackage{array}
\usetikzlibrary{shapes,arrows, positioning}

\def\BibTeX{{\rm B\kern-.05em{\sc i\kern-.025em b}\kern-.08em
    T\kern-.1667em\lower.7ex\hbox{E}\kern-.125emX}}
\begin{document}

\tikzstyle{block} = [rectangle, draw, rounded corners, minimum height=3em]
\tikzstyle{block2} = [rectangle, text width=3.5em, text centered, rounded corners, minimum height=2em]
\tikzstyle{line} = [draw, -latex']

\title{On-site Online Feature Selection for Classification of Switchgear Actuations}

\author{\IEEEauthorblockN{Christina Nicolaou\IEEEauthorrefmark{1}\IEEEauthorrefmark{2}, Ahmad Mansour\IEEEauthorrefmark{1}, Kristof Van Laerhoven\IEEEauthorrefmark{2}}
\IEEEauthorblockA{\IEEEauthorrefmark{1}Corporate Sector Research and Advance Engineering, Robert Bosch GmbH, Renningen, Germany \\
Email: \{christina.nicolaou2, ahmad.mansour\}@de.bosch.com}
\IEEEauthorblockA{\IEEEauthorrefmark{2}Department of Electrical Engineering and Computer Science, University of Siegen, Siegen, Germany, Email: kvl@uni-siegen.de}
}

\maketitle

\begin{abstract}
As connected sensors continue to evolve, interest in low-voltage monitoring solutions is increasing. This also applies in the area of switchgear monitoring, where the detection of switch actions, their differentiation and aging are of fundamental interest. In particular, the universal applicability for various types of construction plays a major role. Methods in which design-specific features are learned in an offline training are therefore less suitable for assessing the condition of switchgears. A new computational efficient method for intelligent online feature selection is presented, which can be used to train a model for the addressed use cases on-site. Process- and design-specific features can be learned locally (e.g. on a sensor system) without the need of prior offline training. The proposed method is evaluated on four datasets of switchgear measurements, which were recorded using microelectromechanical system (MEMS) based sensors (acoustic and vibration). Furthermore, we show that the features selected by our method can be used to track changes in switching processes due to aging effects.  

\end{abstract}


\section{Introduction}
Switchgears are important components of transformer substations and play an indispensable role in the electrical distribution grid by protecting and de-energizing stations in the event of a fault. It is therefore essential to maintain a reliable operation over decades during which the switches are in use and are inevitably affected by aging and wear effects. In Germany, for example, automated switching devices with control sensors at medium-voltage (MV) level are the exception rather than the rule. It is therefore not automatically recorded whether a switching action has been carried out, which, in addition to the switching frequency and the switched current, decisively determines the aging. In contrast, at the high-voltage level, cost-intensive switchgear diagnosis systems record a large number of parameters for condition monitoring. These include among others opening and closing times, drive motor current, spring drive tension and electrical parameters \cite{Nurubeyli2015}. From an economic point of view, extensive condition assessment with cost-intensive monitoring systems is not reasonable for medium-voltage switchgears, since the cost of a monitoring system is restricted by the low cost of the MV switchgear itself \cite{Shi2002}.

In this paper, we show that basic monitoring tasks such as switch identification, switching frequency or aging effects can be detected using cost-effective MEMS-based sensors in combination with vibration and acoustic analysis. In contrast to existing solutions relying on the monitoring of electrical parameters and on limit switches at each breaker to extract the current position of the gear, the underlying idea is to only use one sensor system for monitoring the complete switchgear unit. This greatly simplifies the installation effort, in particular the cabling, and promises to provide a suitable solution for network operators as well as a widespread use. 

Furthermore, to cope with the high diversity of switchgear types and manufacturers, generalizable models for the data processing are required. Typically, records are collected, analyzed offline using application-specific algorithms and then implemented on a gateway or a microcontroller of the sensor system. The repetition of those offline learning cycles for different switchgear construction types leads to a considerable effort and is not feasible for this application. For this reason, we propose an online configuration concept for automatic classification based on intelligent feature selection to differentiate between various switching types and processes (e.g. switching-on and -off). It only needs few labeled training data for a single learning phase that consists of evaluation and online selection of a rather small number of appropriate features for classification using an adapted Silhouette score. This eliminates the need for offline training in the back-end and allows its use on-site with constrained computational power (e.g. on a gateway or embedded on a sensor system). Furthermore, an online assessment of the process or switchgear condition is carried out automatically with the aid of the selected features and the derived model. The proposed method is not only limited to switchgear process classification, but can also be customized for other applications in the machine condition monitoring domain. 

In the following, a short introduction into switchgear and an overview of existing switchgear monitoring solutions is given before further introducing our proposed method for intelligent online feature selection. The approach is evaluated on four datasets of switchgear measurements in a proof-of-concept study.  

\section{Medium-Voltage Switchgear}
\begin{figure}[tb]
	\centerline{\includegraphics[width=0.48\textwidth]{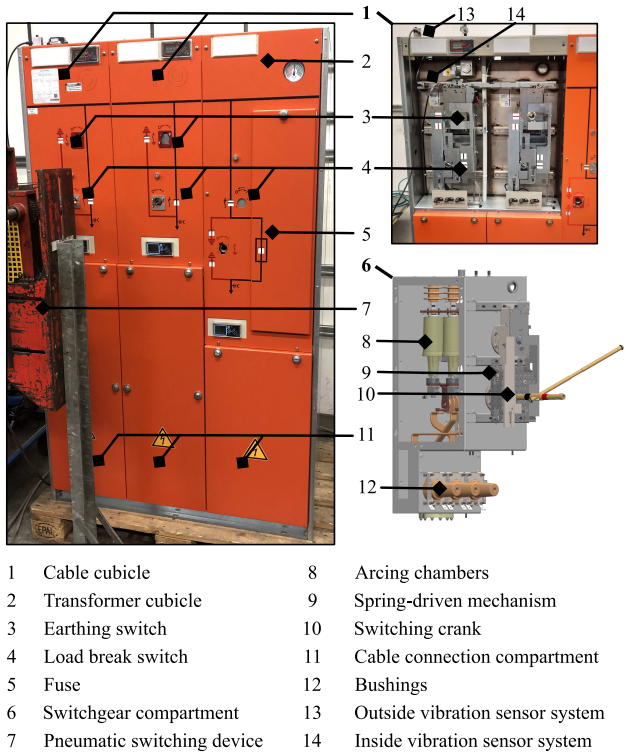}}
	\caption{Example of a triple cubicle switchgear unit from Driescher Wegberg (left) with view into the interior structure of the cable cubicle (top right) and switchgear compartment (bottom right).}
	\label{fig1}
\end{figure}

Medium-voltage switchgear can be classified as open-air installations and encapsulated switchgears (Fig.~\ref{fig1}). The second type is manufactured according to a modular principle from individual metal-encapsulated cubicles (e.g. feed-in cable and transformer outgoing feeder cubicles) consisting of circuit breakers, fuses and switches. A switchgear unit typically contains at least three of them, with the individual cubicles in turn consisting of several compartments. In order to ensure a safe switching operation, spring-driven mechanisms are widely used. The springs are pre-tensioned automatically or manually with a switching crank and release enough energy to complete opening or closing operations independent of the operational speed.

\section{Related Work}
Both vibration and acoustic signals are generated during the opening and closing of switches and can be used for non-intrusive online monitoring tasks. The development of reliable detection algorithms is challenging due to a high variety of switchgear components, an extremely short switching duration (up to a few hundred milliseconds) and the fact that switch operations are mainly done for maintenance reasons few times a year. A considerable amount of work has been done to address these challenges, with the proposed signal processing methods developed and tested mainly for high voltage open-air circuit breakers. Among the methods investigated are contact travel time analysis \cite{Ukil2013}, wavelet analysis \cite{Lee2003}, and a combination of dynamic time warping for time axis alignment, envelope extraction, and correlation analysis \cite{Landry2008}. The last approach compared measured signals with a reference signal of the same switchgear in a healthy condition or to another one from the same type. Amplitude and time deviations were evaluated and compared to an alarm threshold. Furthermore, they showed that each event in the vibration signal can be correlated to the contact position of the breaker. A similar approach was followed in \cite{Hoidalen2005}, which furthermore considered shifts in the frequency content of the signal. All methods used at least one accelerometer per switch. Yang et al. employed a refined approach, where slight changes are found by decomposing the signal into several non-interference modes. For the comparison of signals, a change-sensitive feature was calculated \cite{Yang2019}. 

While these methods can detect the exact cause of a mechanical failure (e.g. poor lubrication, contact damage), they require high manual effort to adjust the thresholds for different types of switchgear and detailed expert knowledge to link vibration events to certain defects. In addition, the comparison of the signals is highly dependent on the positioning of the accelerometers. However, this effort and the necessary precise sensors with high bandwidth are not justified at MV level, which is why self-learning algorithms and the use of MEMS-systems are examined with the following method.

\section{Intelligent Online Feature Selection}
\begin{figure} [b]
	\footnotesize
	\begin{minipage}{\linewidth}
		\begin{center}
			\begin{tikzpicture}[node distance = 0.5cm and 0.3cm, scale=1, transform shape]
			\node(training) [block2] {\textbf{Training:}};
			\node [block, text width=5.5em, right = of training] (init) {Labeled data of scenarios};
			\node [block, text width=4em, right = of init] (features) {Feature extraction};
			\node [block, text width=6em, right = of features] (fs) {Feature selec-tion via $fc$};
			\node [block, text width=4.5em, right = of fs] (model) {Set model};
			\node(inference) [block2, below = of training] {\textbf{Inference:}};
			\node [block, text width=4em, right = of inference] (signal) {Incoming data};
			\node [block,  text width=6.5em, right = of signal] (filtfeat) {Extract selected features};
			\node [block, text width=4.5em, right = of filtfeat] (process) {Scenario estimation};

			\path [line] (init) -- (features);
			\path [line] (features) -- (fs);
			\path [line] (fs) -- (model);

			\path [line] (signal) -- (filtfeat);
			\path [line] (filtfeat) -- (process);
			\end{tikzpicture}
		\end{center}
	\end{minipage}
	\caption{Configuration principle of intelligent online feature selection with training and inference phases.}
	\label{flow}
\end{figure}
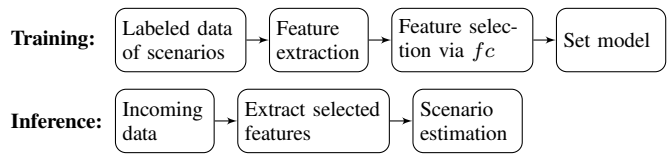

The proposed intelligent online feature selection algorithm is summarized in Fig.~\ref{flow}. In the first step, the algorithm is provided with labeled training data $x_T$ on the processes/scenarios to be distinguished. For representing the time series data, features in time and frequency domains are extracted to map the data $x_T$ in the feature space, where the characteristics of different processes $k$ are better distinguishable. Appropriate features must be specified by the user in advance, whereby this provides the possibility to incorporate expert as well as physical knowledge. Furthermore, there are no restrictions on the number of features. Packages extracting a high amount of different features exist in Python (tsfresh \cite{Christ2018}, 794 extracted features) and MATLAB (hctsa \cite{Fulcher2017}, 7700 extracted features). Those packages also use highly parallelized feature selection algorithms based on statistical hypothesis tests, which require high computational power and are therefore not suitable for our use case.

\normalsize

\begin{figure}[bp]
	\centerline{\includegraphics[width=0.50\textwidth]{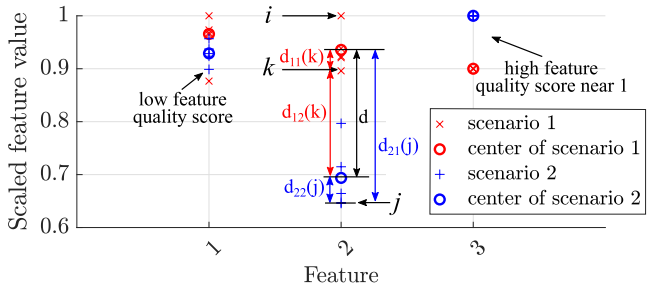}}
	\caption{Example of three different scaled features and the distances for point j and k for the feature quality score calculation.}
	\label{fig3}
\end{figure}

For classification or clustering problems, the generalizability beyond the training set is a major concern, which is why the significance of extracted features is of high relevance and the selection of too many irrelevant features needs to be avoided. For the feature selection, a combination of clustering and classification is used to identify the relevant features. To achieve this, the training data labels are compared with the clustering quality, which is determined using a modified Silhouette analysis. The Silhouette score $S_i = \frac{b_i-a_i}{\max(a_i, b_i)}$ is a measure on how similar a feature point $i$ is to points of its own cluster (cohesion) compared to points of other clusters  (separation) \cite{Kaufman1990}. For the similarity, the average intra-cluster distance $a_i$ and inter-cluster distance $b_i$ are calculated. The disadvantage of this approach is the high computational effort for calculating the distances $a_i$ and $b_i$ between all points, which scales with the number of training data (number of features and number of samples). Our adaption instead calculates a feature quality score ($fc$) out of the distance between the features and their own cluster center $c_i$ and the nearest cluster center $c_k$ as follows:
\begin{equation}
	fc =\frac{1}{N}\sum_{i=1}^{N}\frac{d_{ik}-d_{ii}}{\max(d_{ik}, d_{ii})},
	\label{eq1}
\end{equation}
whereby $N$ is the number of different processes, $d_{ii}$ is the average distance of feature values of one cluster to $c_i$ and $d_{ik}$ is the average distance of those features to the nearest cluster center $c_k$ (Fig.~\ref{fig3}). A good classification is given for a feature quality score near 1.
In the inference phase, the data are continuously collected and then automatically analyzed. If a switch operation is detected, the selected features are calculated and an estimation of the detected process from the model based on the classification probability is given to the user. This probability is calculated based on the feature distance to the centers as shown in Fig.~\ref{fig3}. Point $i$ is assigned to scenario/process 1 with a probability of 100~\%, and point $j$ correspondingly to scenario 2. Point $k$, laying between the clusters, is assigned to scenario 1 with a probability of $P=1-\frac{d_{11}}{d}$ ($d$ is the distance between the centers).

\section{Proof-Of-Concept Study}
The benefits of intelligent online feature selection are studied here for the differentiation of switching processes on four datasets. We implement a switchgear actuation detection to detect differentiable processes and automatically label them for the training phase of the proposed method, illustrate the results and show how the selected features can be used for aging detection.

\subsection{Data} \label{data}
This study's data were recorded at gas-isolated (SF6) switchgear using a MEMS-based stereo microphone (\textit{MBSM}) with a sampling frequency of 48~kHz and two MEMS vibration sensor systems (\textit{VSS}) consisting of two accelerometers each. These are used to cover a high bandwidth as well as a good resolution in the low frequency range. The first accelerometer ($A1$) has a single degree of freedom and a sampling rate of 62.5~kHz, while the second one ($A2$) is a three-axis sensor with a sampling rate of 2~kHz. The microphone was installed in a distance of approximately two meters behind the switchgear units, one vibration sensor (\textit{VSS1}) was screwed inside the cubicle next to the spring-drive mechanism, whereas the second system (\textit{VSS2}) was mounted on top of the corresponding cubicle (see Fig.~\ref{fig1}). All sensor systems feature integrated A/D conversion, amplification and correction factors, thus providing processed digital values. Experiments were conducted on different types of switches and in different cubicles. Details of the datasets DS1 to DS4 are listed in Table \ref{tab1}. Both the switching of the units as well as the measurements were automated. The data recording was triggered by limit switches, which were turned after each opening and closing operation. As the limit switches tend to bounce, the number of measurement files in one dataset is not consistent over all sensors.   
\begin{table}[bp]
	\caption{Datasets and parameter variations}
	\begin{center}
		\begin{tabular}{|m{0.1\linewidth}|m{0.4\linewidth}|m{0.1\linewidth}|m{0.1\linewidth}<{\centering}|}
			\hline
			\textbf{Dataset}&\textbf{Switchgear type}&\textbf{Sensor}&\textbf{Records}\\
			\hline
			\multirow{3}{*}{DS1} & \multirow{3}{0.9\linewidth}{load break switch in cable cubicle}  & VSS1 & 585\\ \cline{3-4} &  & VSS2 & 948 \\ \cline{3-4}& & MBSM & 1118 \\
			\hline
			\multirow{3}{*}{DS2} & \multirow{3}{0.9\linewidth}{load break switch in cable cubicle}  & VSS1 & 306\\ \cline{3-4}&  & VSS2 & 27 \\ \cline{3-4}& & MBSM & 279 \\
			\hline
			\multirow{3}{*}{DS3} & \multirow{3}{0.9\linewidth}{load break switch in transformer cubicle}  & VSS1 & 3295\\ \cline{3-4}&  & VSS2 & 3299 \\ \cline{3-4}& & MBSM & 3187 \\
			\hline
			\multirow{3}{*}{DS4} & \multirow{3}{0.9\linewidth}{vacuum circuit breaker}  & VSS1 & 3830\\ \cline{3-4} &  & VSS2 & 533 \\ \cline{3-4} & & MBSM & 3831 \\
			\hline
		\end{tabular}
		\label{tab1}
	\end{center}
\end{table}

\subsection{Actuation Detection} \label{action}
\begin{figure}[tbp]
	\centerline{\includegraphics[width=0.48\textwidth]{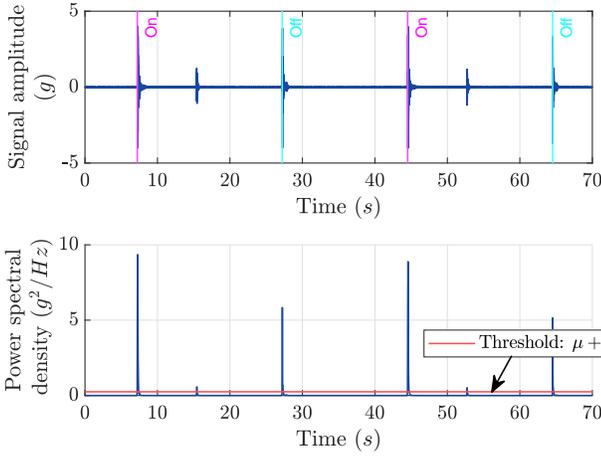}}
	\caption{Vibration signal from DS3 (above) and its power spectral density computed over intervals of 33~ms (below). A switchgear operation is detected when exceeding the threshold, whereby no new operation can be detected during the switching duration.}
	\label{fig2}
\end{figure}

Acoustic and vibration signals from switching operations are characterized by their short duration as well as their high amplitudes as shown in Fig.~\ref{fig2}. With a signal-to-noise ratio (SNR) of around 20~dB over all measurements, they can be distinguished from background noise very well. To detect switchgear operations for further classification of different processes, a threshold recognition is implemented. In order to reduce the amount of data and increase the robustness against outliers, the raw data are compressed by computing the sum of power spectral density (PSD) estimates in intervals of equal lengths (Fig.~\ref{fig2}). From the average and standard deviation of the PSD values of training data, the detection threshold is calculated. When switching a circuit breaker as in Fig.~\ref{fig2}, a spring is pre-tensioned to guarantee that the switch-off process can be carried out safely. When this spring snaps into place, a second peak is detected. To ensure that only one process is identified, new start points within the average switching time of the training data are discarded. 

For DS1-DS4, this leads to a detection accuracy of $>99\%$ for all sensors, whereby the method is robust against usual background noises (e.g. speaking, loud coughing) that were captured during the measurements.

\subsection{Differentiation of Switchgear Processes} \label{process}
In addition to the basic detection, the determination of the number and frequency of switching is of great interest. When having more than one switch installed, the simple counting of detected operations is no longer sufficient and a signal-based differentiation of processes (e.g. on/off, switchgear types) becomes necessary. For the first step, the sensor is installed in its final position and the different processes that need to be classified are executed several times. For a realistic practicability, five recurrences have been selected for each process. Differentiable processes can be the switching of different breaker types, new and old switchgear, as well as switching-on and -off, which is the focus here. For this use case, the labeling of the training data is done automatically by first detecting the actuations with the above algorithm and then alternately labeling it with "on" and "off" (Fig.~\ref{fig2}). In a second step, the training data $x_T$ are low pass filtered with different cut-off frequencies $f_j$ as the damped oscillation of the springs lies mainly in the low-frequency range (\textless10~kHz \cite{Lee2003}). Features calculation is executed for each of the filtered signals $X_{T,f_j}$. In this study, 21 features listed in Table \ref{tab2} were extracted and afterwards rescaled (e.g. z-score) for data normalization. The best five features were selected with the proposed method and used for the inference.

\begin{table}[tbp]
	\caption{Time and frequency domain features}
	\begin{center}
		\begin{tabular}{|m{0.25\linewidth}|m{0.6\linewidth}|}
			\hline
			\textbf{Domain}&\textbf{Features}\\
			\hline
			Frequency \& Time & mean, variance, skew, kurtosis, power, flatness\\
			\hline
			Time & root mean square, absolute mean, maximum, minimum, dynamic range, crest factor\\
			\hline
			Frequency & spectral centroid, median frequency, dominant frequency\\		
			\hline
		\end{tabular}
		\label{tab2}
	\end{center}
\end{table}

\subsection{Results} \label{results}
The average accuracy per sensor over all selected features and axes is shown in Table \ref{tab3} (\textit{Selected features}). For the differentiation between switching-on and -off, the microphone leads to the best results over all datasets with an accuracy of 96.2~\%, whereby the side further away from the switchgear unit results in an accuracy of 98.6~\%. The reason for this phenomenon is the fact that the sensors exceed their defined measuring range and start to clip if being too close to the tested unit due to the large forces released by the switching process. This also leads to artifacts in the measurements. The same can be seen in the results of the two vibration sensor systems. The outside installed system \textit{VSS2} leads to a better detection compared to the sensors installed near the switchgear drive. As the measurements show a damped oscillation frequency, features from the frequency domain are primarily selected. By far, the most frequently selected feature is the dominant frequency in 35~\% of the cases with an average accuracy over all datasets of 85~\% (all sensors included). The extracted dominant frequencies of the two processes are mostly located below 1~kHz. This supports the approach of filtering the signal before extracting the features and also explains why the accelerometers \textit{A1} with a greater resolution in the low frequency range performs better. In contrast to the selected features, the axis orientation of \textit{A1} plays a minor role for the accuracy. Trigger problems and therefore inconsistent truncated measurements result in a lower accuracy in DS2. To reduce the dependency on individual features for the scenario estimation, a majority vote of the result from the five extracted features is implemented. Three different versions (equally weighted, weighted with feature quality score, weighted with scenario estimation probability) are tested, which all lead to a similar accuracy (0.5\% deviation). Table \ref{tab3} (\textit{Majority vote}) shows the equally weighted feature results, which, in most cases, improves the overall accuracy compared to the first column. Learned features from DS1 were also tested on the inference data of DS2, which lead to worse results. Features learned for one sensor position cannot easily be transferred to other positions since the signal characteristics differ if the setup changes. This shows the necessity of our approach to give an opportunity for on-site learning. 

\subsection{Trend Detection for Aging Effects} \label{age}
\begin{figure}[tp]
	\centerline{\includegraphics[width=0.5\textwidth]{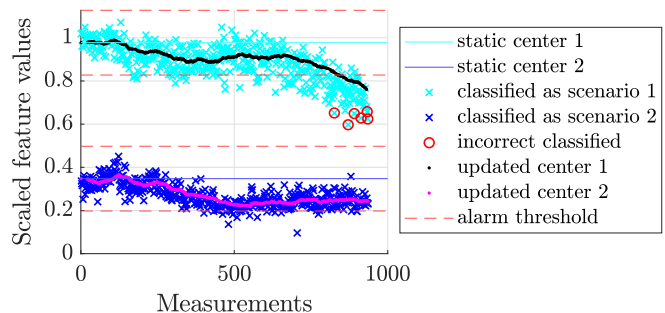}}
	\caption{Classification of the VSS2 A2 measurements of DS1. Updated centers allow a better accuracy over time and can track changes. By exceeding one of the thresholds, an alarm is triggered. 
	\vspace{-5mm}}
	\label{fig4}
\end{figure}

In addition to the differentiation of various processes, aging processes can also be mapped with the help of the selected features. At the end of the measurements of DS1, there was a failure of the spring-based switching mechanism of the load switch. Fig.~\ref{fig4} shows the change in one of the selected features for \textit{VSS2 A2}. As data from broken switches are not always available, the above method cannot always be used to differentiate between new and used switchgear to track changes. An alternative is the change tracking of the already selected features by continuously updating the calculated cluster centers to track their development. The update is done with the exponential weighted moving average (EWMA)
\begin{equation}
	c_{ij} = \alpha * f_{ij} + (1-\alpha)*c_{ij-1},
	\label{eq2}
\end{equation}
whereby $c_{ij}$ describes the center of a scenario $i$, $f_{ij}$ is the corresponding feature at time $j$ and $\alpha\in[0,1]$ is a weighting factor to determine how much influence the new feature $f_{ij}$  has on the old center $c_{ij-1}$ . In our implementation, only features in the $\mu\pm3\sigma$-range of the center were considered for the calculation and $\alpha=0.05$ was used. This approach leads to an improvement of the accuracy as shown in \textit{Updated centers} of Table \ref{tab3}. The change between the current value of the center and the beginning one is a good indicator for the aging of switchgear. By using definable alarm thresholds, an alarm can be triggered in case of a major change (Fig.~\ref{fig4}). 
\begin{table}[tbp]
	\caption{Accuracy of the online feature selection for DS1-DS4}
	\begin{center}
		\begin{tabular}{|m{0.09\linewidth}|m{0.14\linewidth}|m{0.12\linewidth}<{\centering}|m{0.12\linewidth}<{\centering}|m{0.12\linewidth}<{\centering}|}
			\hline \multirow{2}{*}{\textbf{Dataset}} & \multirow{2}{*}{\textbf{Sensor}} & \multicolumn{3}{c|}{\textbf{Accuracy in [\%]}} \\ \cline{3-5} & & \textbf{Selected features} & \textbf{Majority vote} & \textbf{Updated centers}\\
			\hline
			\multirow{5}{*}{DS1} & VSS1 A1 & 70.8 & 75.5 & 91.4\\
			\cline{2-5} & VSS1 A2 & 81.7 &  89.1 &  96.9\\
			\cline{2-5} & VSS2 A1 & 85.9 & 85.9 & 85.9\\ 
			\cline{2-5} & VSS2 A2 & 89.0 &  95.3 & 99.8 \\
			\cline{2-5} & MBSM & 98.7 & 98.9 & 99.1\\
			\hline
			\multirow{5}{*}{DS2} & VSS1 A1 & 57.1 &  57.1 & 57.1\\
			\cline{2-5} & VSS1 A2 & 66.0 &  68.4 & 65.7 \\
			\cline{2-5} & VSS2 A1 & 78.6 &  78.6 & 78.6\\
			\cline{2-5} & VSS2 A2 & 76.7 &  85.4 & 85.4 \\
			\cline{2-5} & MBSM & 89.0 &  93.6 & 94.4\\
			\hline
			\multirow{5}{*}{DS3} & VSS1 A1 & 98.5 & 98.5 & 98.6\\
			\cline{2-5} & VSS1 A2 & 99.0 & 99.3 & 99.3\\
			\cline{2-5} & VSS2 A1 & 100  & 100 & 99.9 \\ 
			\cline{2-5} & VSS2 A2 & 96.1 & 97.6 & 97.6\\
			\cline{2-5} & MBSM & 94.8 & 94.8 & 99.7\\
			\hline
			\multirow{5}{*}{DS4} & VSS1 A1 & 93.8  & 95.2 & 100\\
			\cline{2-5} & VSS1 A2 & 91.6 & 98.0 & 100 \\
			\cline{2-5} & VSS2 A1 & 97.8 & 97.9 & 99.8\\ 
			\cline{2-5} & VSS2 A2 & 97.8 & 98.1 & 98.1\\
			\cline{2-5} & MBSM & 97.1 & 98.9 & 98.8\\
			\hline	
		\end{tabular}
		\label{tab3}
	\end{center}
\end{table}

\section{Conclusions}
In this paper, we present an approach for on-site configuration for classification based on an online intelligent feature selection algorithm, which eliminates the need for time-consuming, application-specific offline training. We show that, with this approach, switchgear actuations that are automatically detected via a threshold can be classified reliably for different scenarios. The suitability of MEMS-sensors is evaluated and best practices for the installation are given. Furthermore, the selected features can be used to detect trends in the data for switchgear aging. An update of the cluster centers further improves the classification quality by 6~\% on average in accuracy.
Further investigations include extending the approach to multi-class classification and evaluating the usability of only one sensor system, both for monitoring a complete switchgear unit.

\section*{Acknowledgment}
The project MAKSIM, on which this report is based, is funded by the Federal Ministry for Economic Affairs and Energy under the funding code 0350035B. The authors are responsible for the content of this publication. The authors would like to thank the Fritz Driescher KG for the cooperation and for providing access to their switchgears. Thanks goes especially to Mr.~Bernards and Mr.~Goertz for their help and support in the study and tests.

\bibliography{switchgear}
\bibliographystyle{IEEEtran}

\vspace{12pt}
\end{document}